\documentclass{elsart}

 \usepackage{graphicx}
\usepackage{amssymb}

\begin{document}

\begin{frontmatter}



%\title{}

\title{Photomultiplier Tubes in the MiniBooNE Experiment}

\author[FNAL]{S. J. Brice,}  
\author[Columbia]{L. Bugel,}
\author[Columbia]{J. M. Conrad,}
\author[Yale]{B. Fleming,}
\author[Columbia]{L. Gladstone,\corauthref{cor}}
\corauth[cor]{Corresponding author. Tel: 917-647-5099; fax: 212-854-3379}
\ead{leg2102@columbia.edu}
\author[WIL]{E. Hawker,}
\author[Michigan]{P. Killewald,}
\author[SLAC]{J. May,}
\author[LANL]{S. McKenney,}
\author[Paul's]{P. Nienaber,}
\author[Michigan]{B. Roe,}
\author[LANL]{V. Sandberg,}
\author[ERAU]{D. Smith,}
\author[Michigan]{M. Wysocki}

\address[FNAL]{Fermi National Accelerator Laboratory, %PO Box 500 
Batavia IL 60510} 
\address[Columbia]{Columbia University, Pupin Laboratories, 538 W. 120th St., New York NY 10027} 
\address[Yale]{Yale University, New Haven CT 06520} 
\address[WIL]{Western Illinois University, Currens Hall, Macomb IL 61455} 
\address[Michigan]{University of Michigan, Department of Physics, 
Ann Arbor MI 48109} 
\address[SLAC]{Stanford Linear Accelerator Center, 
Menlo Park CA 94025}%This is Justin's personal address
\address[LANL]{Los Alamos National Laboratory, %P.O. Box 1663,
Los Alamos NM 87545}
\address[Paul's]{Saint Mary's University of Minnesota, Winona MN 55987} 
%\address[Princeton]{Princeton University, Joseph Henry Laboratories,
%Princeton NJ 08544} 
\address[ERAU]{Embry-Riddle Aeronautical University, Prescott AZ 86301}

\date{\today}


\address{}

\begin{abstract}
  The detector for the MiniBooNE~\cite{booneproposal} experiment at
  the Fermi National Accelerator Laboratory employs 1520 8 inch
  Hamamatsu models R1408 and R5912 photomultiplier tubes with
  custom-designed bases.  Tests were performed to determine the dark
  rate, charge and timing resolutions, double-pulsing rate,
  and desired operating voltage for each tube, so that the tubes could
  be sorted for optimal placement in the detector.  Seven phototubes
  were tested to find the angular dependence of their response. After
  the Super-K phototube implosion accident, an analysis was performed
  to determine the risk of a similar accident with MiniBooNE.
\end{abstract}

\end{frontmatter}

\section{Introduction}

\begin{figure}
%[htb bbllx=65bp,bblly=75bp,bburx=546bp,bbury=716bp,width=4.in]
\centering
\includegraphics[height=350pt]{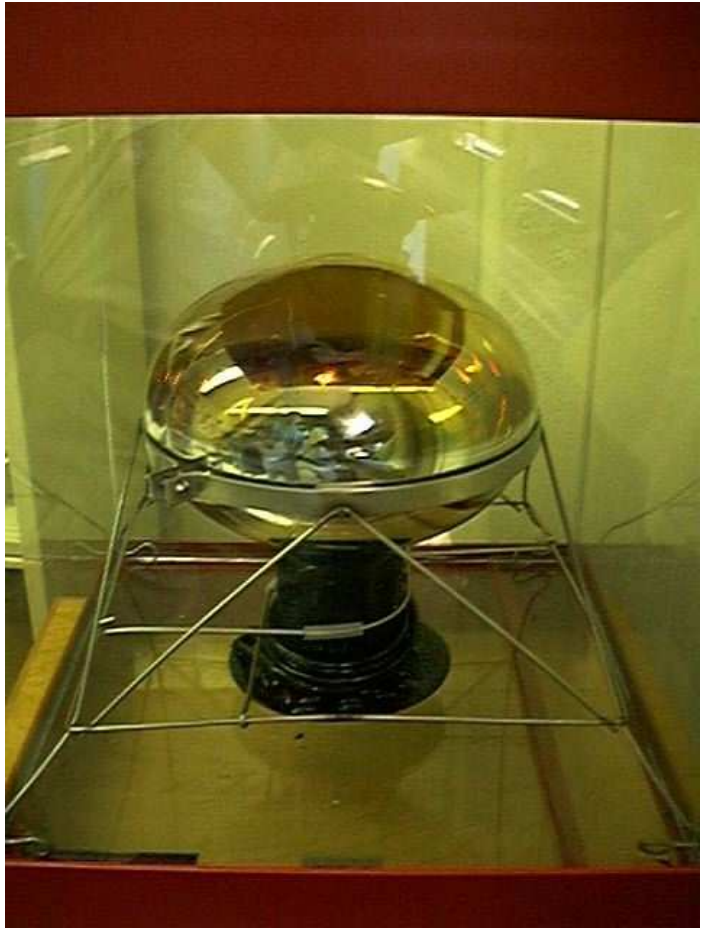}
\caption{A typical phototube used in the MiniBooNE detector: an 8 inch
  R1408 mounted on its wire frame. The base is coated with Masterbond
  (as described in section \ref{pgh: masterbond}) before the PMT is
  installed into the detector.}
  \label{lsnd_pmt}
\end{figure}

The MiniBooNE experiment~\cite{booneproposal} is a $\nu_\mu
\rightarrow \nu_e$ oscillation search at Fermi National Accelerator
Laboratory designed to confirm or rule out the LSND
signal~\cite{Athanassopoulos:1996jb}. The Fermilab Booster accelerates
protons to 8 GeV; these protons strike a beryllium target, generating
mesons which decay to produce the MiniBooNE $\nu_\mu$ beam.  The
neutrinos interact in a 12 m diameter sphere filled with mineral oil of food-grade purity, and the \v{C}erenkov light from charged particles
produced in these interactions is detected by 8 inch Hamamatsu~\cite{hamamatsu website}
photomultiplier tubes (PMTs) lining the sphere. There are 1280 PMTs\ in the light-tight inner signal region: 956 are Hamamatsu model R1408, inherited from the LSND experiment; the
remainder are Hamamatsu model R5912. 
Light produced in the outer concentric veto region is detected by an additional 240 R5912 PMTs.
A typical tube is pictured in Figure \ref{lsnd_pmt}.
Light signatures recorded by the PMTs are used to reconstruct events
within the detector. Leptons from charged
current neutrino interactions are of particular interest, since these leptons tag the incoming
neutrino flavor.

It is important to sufficiently understand  the operating characteristics of the PMTs, since they are the only active detector element in MiniBooNE. MiniBooNE uses a Monte Carlo simulation to understand the detector's response to neutrino interactions. 
This program simulates photons
traveling through the detector from creation to detection, and models
all intermediate processes. 
 The geometry of all detector elements is
coded into the simulation, including that of the PMTs.

%The PMTs are the only active detector elements in MiniBooNE; it is therefore important to measure their operating characteristics so that the sensitivity and accuracy of the MiniBooNE detector can be understood and optimized.

Photons striking the photocathode in a PMT produce photoelectrons (PEs),
whose number is amplified via a dynode chain; the magnitude of the
current so produced is proportional to the number of incident photons
whose energy is above a certain threshold.  A specific operating
voltage must be found for each PMT so that all tubes have the proper
gain. Gain is defined here as observed output charge (divided by charge per electron) per input photoelectron.  The charge and timing
information recorded from each PMT can then be used to reconstruct the
events off-line.  Ideally, this information, along with photocathode efficiency as determined by Hamamatsu, specifies exactly when and
how many photons hit the PMT. 
Each PMT is tested in several ways to determine: basic functionality, timing resolution, charge resolution, dark rate and operating voltage.
The operating voltage was chosen so that the gain is closest to the desired $1.6 \times10^{7}$ signal electrons per photoelectron.

In addition to these global PMT tests, specialty tests were done on
seven PMTs. In order for the Monte Carlo simulation to properly model the
behavior of the PMTs, it is necessary to correctly include their
efficiency as a function of incident photon angle. The dominant effect
comes simply from the solid angle subtended by a PMT as a function of
angle, and this is handled by coding an approximate PMT shape into
the simulation. Any remaining efficiency function must be measured and
entered into the code. The measurement, however, will itself contain the
solid angle effects; these must be factored out of the measured
angular efficiency to avoid double-counting. A thorough understanding
of the angle-dependent response, time and charge resolution, and
individual variations of the photomultiplier tubes will help
accurately identify neutrino events within the MiniBooNE detector.

This paper discusses the following topics. In Section \ref{sec: technical design of the pmts}, the technical specifications of the phototubes are described, along with the custom bases that MiniBooNE used. In Section \ref{sec:global  testing}, the tests which were performed on every tube, with their results, are described. In Section \ref{sec:angular testing}, the angular tests which were performed on 7 tubes are described, along with their results and a comparison of these results to other angular tests. Section \ref{sec:implosion risk studies} describes the studies of MiniBooNE's risk of an implosion chain reaction.

\section{Technical Design of the PMTs}
\label{sec: technical design of the pmts}
Phototubes in the MiniBooNE detector satisfy the following technical
requirements.  Their maximum dark rates lie below the response rate expected from cosmic rays. Their gains are high, around $10^7$, to be adequately sensitive. Their responses are
reliably fast, and have a narrow enough pulse width to minimize response pileup. Each of these characteristics was tested; these tests are described in subsequent sections. 

\subsection{PMT Specifications from Hamamatsu and custom base design}

The two models (R5912 and R1408) of phototubes used in MiniBooNE differ primarily in their dynode structures and number of stages, and are similar in all the characteristics described above. 
 
\begin{figure}% [htbp bbllx=470bp,bblly=262bp,bburx=756bp,bbury=445bp]
\begin{center}
\includegraphics[bb=1 1 294 190 width=\textwidth]{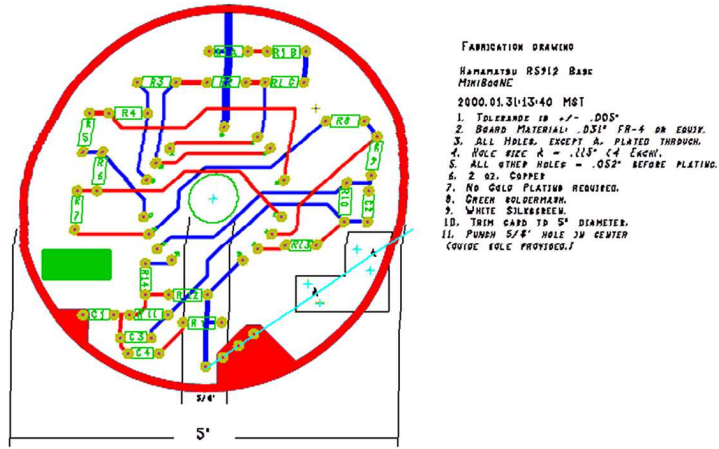}
\caption{The printed circuit artwork and layout for the custom bases of the R5912 phototubes. The design is a two-sided card made from epoxy fiberglass (G10). For a circuit schematic, see Figure \ref{hamamatsu_drawing_dynode}. }
\label{baseart}
\end{center}
\end{figure}

The R5912 is a ten-dynode-stage photomultiplier tube with an 8 inch
hemispherical photocathode. The base supplies the high voltages for
the grids, dynodes, and anode, and provides the back termination in
the anode circuit. The voltage divider chain is entirely passive, and
is fabricated with carbon-film resistors and ceramic-disk capacitors.
A single coaxial cable is used to provide high voltage (on the order
of 1500V) and to connect the anode signal to the readout.  The anode
circuit is back terminated in 50 $\Omega$ and balanced to compensate
for the distortion caused by the capacitive coupling in the
termination circuit. The high voltage taper is the one recommended by
Hamamatsu (ca. 1996) for gain. A  drawing of the tube dome
and a diagram of the dynode circuitry from the Hamamatsu data sheets are included in the appendix as Figure \ref{hamamatsu_drawing_dynode}. Figure \ref{baseart} shows the printed circuitboard layout for the R5912 custom bases.

\begin{figure}[htbp]
\begin{center}
\includegraphics[width=.8\textwidth, bb=190 180 570 650]{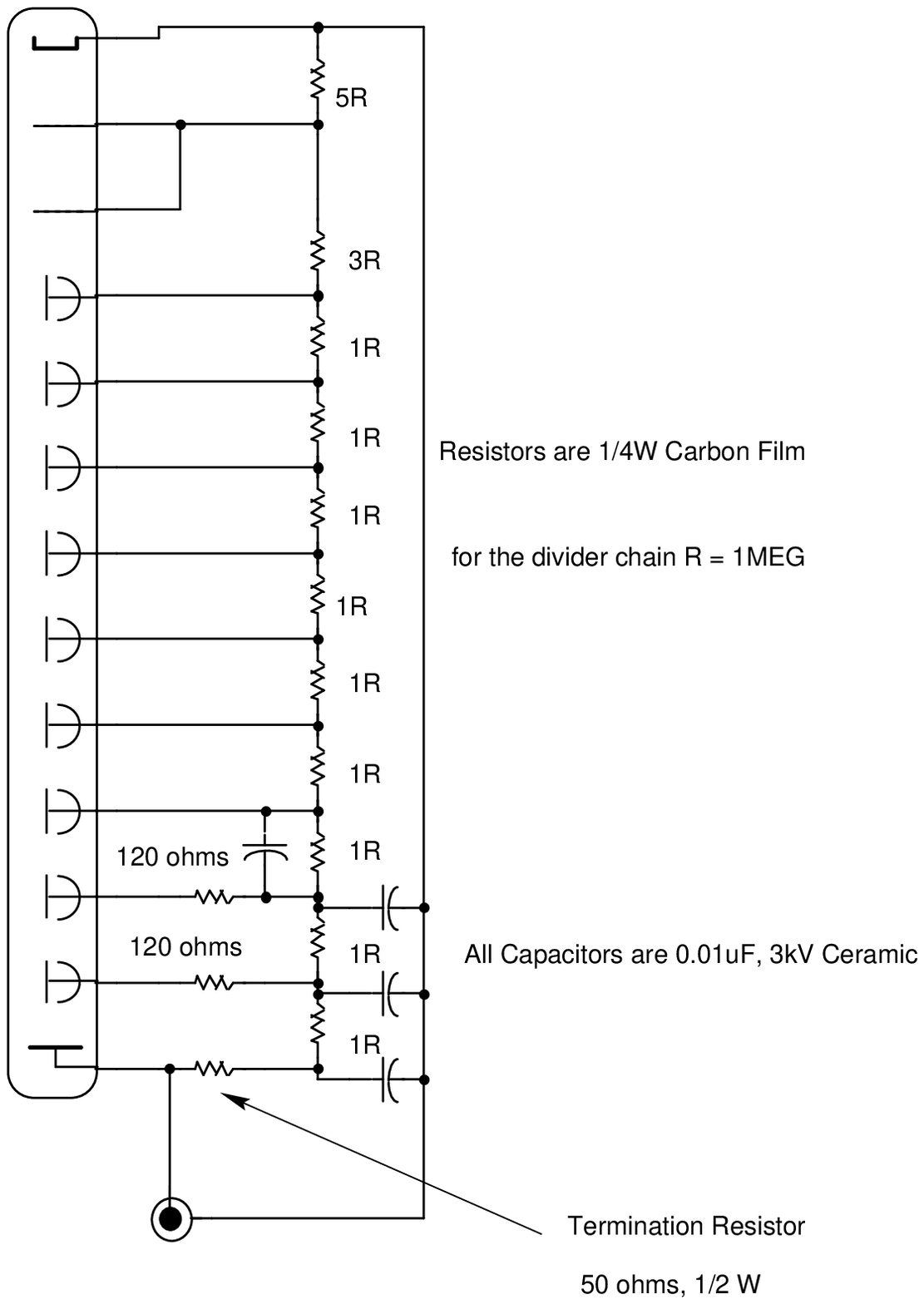}
\caption{Schematic diagram of the base used for the R1408 photomultiplier tube}
\label{r1408base}
\end{center}
\end{figure}

 The R1408 is a nine-dynode-stage photomultiplier tube. Hamamatsu upgraded it to the R5912 at least 10 years before this publication. The circuit diagram for the base (designed for the LSND experiment) is shown in Figure \ref{r1408base}.

The detector design requires that the tubes, bases, and cables be
immersed in oil for several years without adversely affecting the
phototubes or the oil.  To accomplish this, the bases and necks of all
MiniBooNE PMTs were coated with the two-component epoxy EP21LV from
Masterbond \cite{masterbond}, which is impervious to and insoluble in oil. For the R1408
tubes recycled from LSND, the Masterbond was applied on top of the
existing Hysol \cite{hysol} coating. \label{pgh: masterbond}

\vspace{16pt}

%\begin{figure}[htbp]
%%\centering
%\begin{centering}
%\includegraphics{pmtspice}
%\caption{Model circuitry for the anode waveform, cable driving circuit, and dynode chain with anode, developed for designing bases}
%\label{pmtspice}
%\end{centering}
%\end{figure}

\section{Global Testing}\label{sec:global testing}%~\cite{Fleming:2002mv}}

The method of global testing for the PMTs has been described in 
detail elsewhere~\cite{Fleming:2002mv}; a synopsis of
that report is provided here.

The purpose of global testing was to select an operating voltage which
resulted in a similar response from each PMT, regardless of model (R1408
or R5912).  The selected gain was $1.6 \times $10$^7$ electrons per
photoelectron (${\rm PE}$).

The testing process served to determine five characteristic data
about each PMT: dark rate, time jitter, charge resolution,
double-pulsing rate, and pulse shape.  These data were used to sort
the tubes into five categories, from best to worst in timing and
charge resolution.  Categories 4 and 5 had similar resolutions, and
were divided between the two categories by dark rate.  These
classifications were used to distribute tubes ranked 1 through 4 (with about 320 in each rank)
evenly throughout the tank.  Category 5 tubes were used in the veto,
where timing and charge resolution is less crucial but low dark rate
is important. \label{pgh:categories}

\subsection{Global testing: apparatus}\label{sec:global apparatus}

\begin{figure}[htbp]
\begin{center}
\includegraphics[width=5.0in]{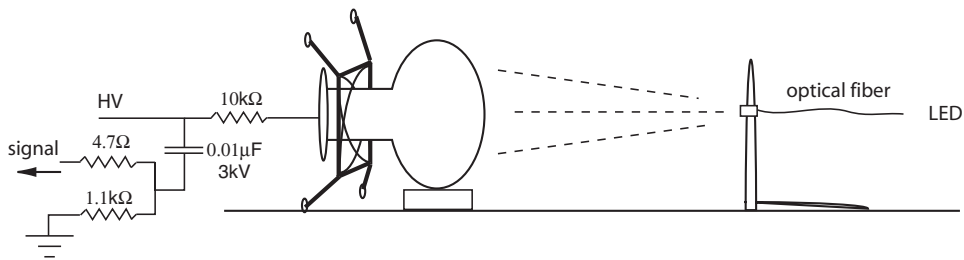}
\end{center}
\caption[test_setup]{The test setup used to measure
  the response of PMTs at low light levels (not to scale).  The pulse
  is picked off the high voltage line shown in the left side of the
  figure.}
  \label{test_setup}
\end{figure}

Figure \ref{test_setup} shows a schematic of a single tube in the test
setup.  The full setup accommodated 30 R1408 PMTs and 16 R5912
 PMTs.  Tubes were dark adapted for 12-24 hours.  After this,
tube responses were recorded using an automated VXI readout system. Fiber optic cables located about eight inches from
the face of each tube transmitted a steady
stream of pulses from an LED pulser.  A diagram of a signal tube setup is shown in Figure \ref{test_setup}.

\begin{figure}[tbp]
\centering
\includegraphics[width=1.1\textwidth]{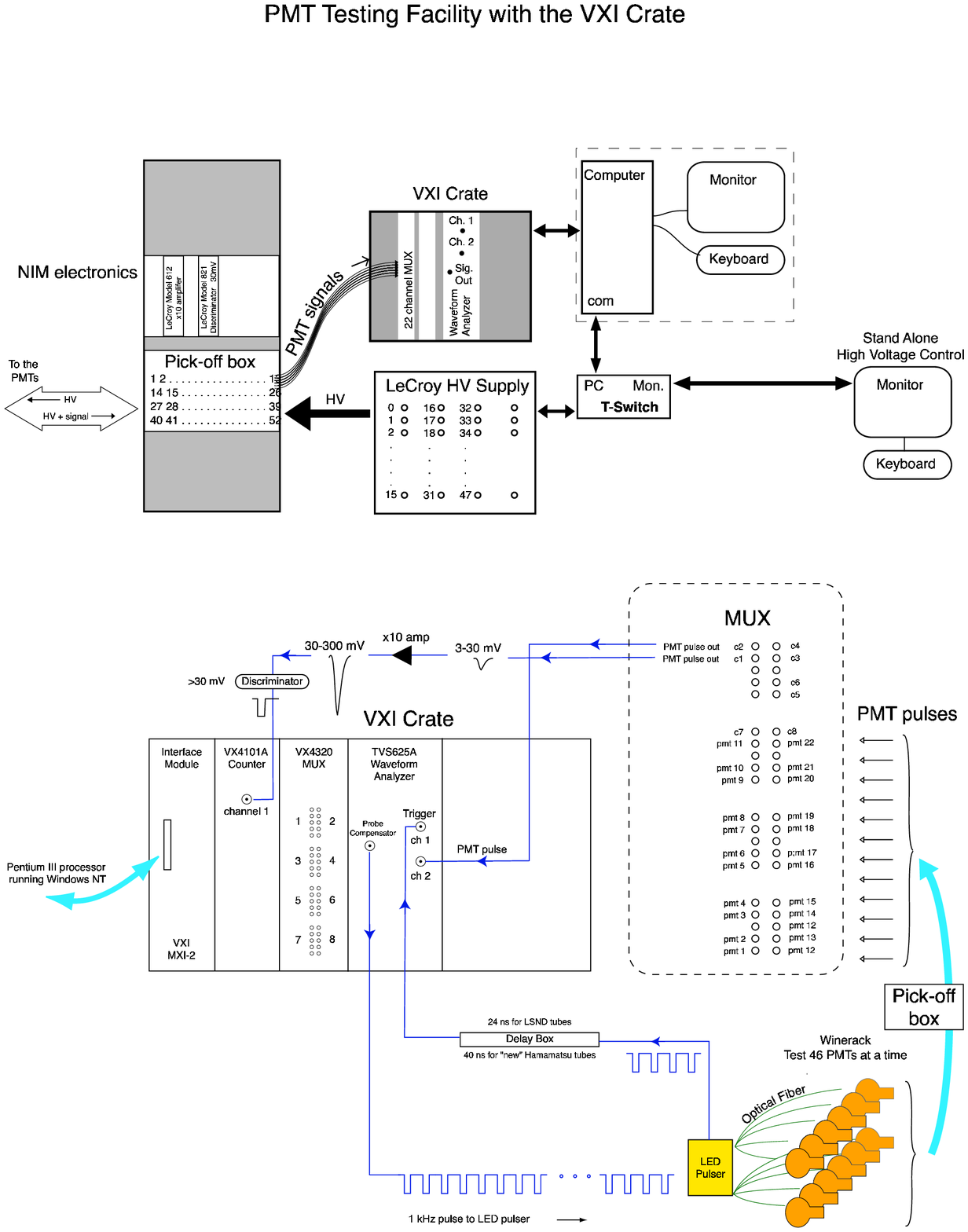} 
  \caption{A schematic of the PMT testing facility data acquisition system. The PMT responses to an LED pulser are aquired from the HV input and analyzed in the VXI crate (from ref.~\cite{Fleming:2002mv}).}
\label{fig:VXI}
\end{figure}

\begin{figure}[ht]
\begin{center}
\includegraphics[width=4.95in]{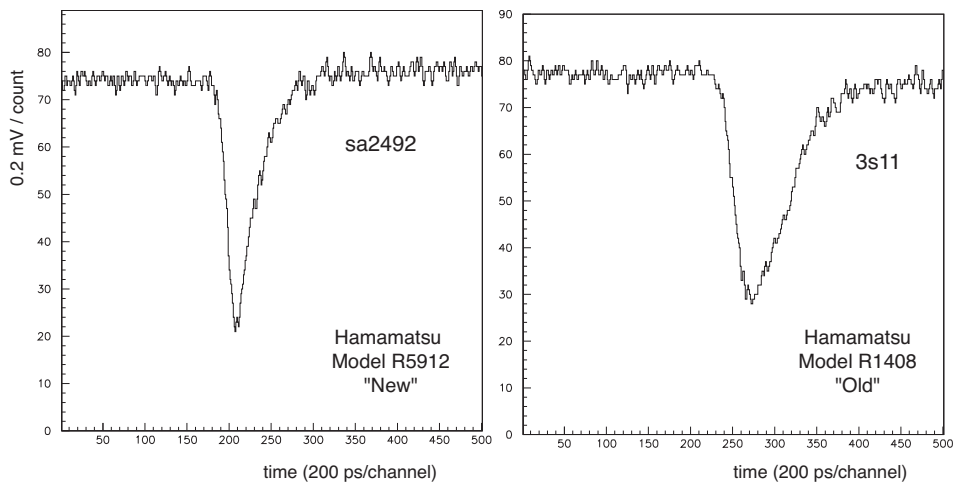}
\end{center}
\caption[waveform]{Single
photoelectron waveforms from the R5912 (left) and
R1408 PMTs.  Typical pulse widths were
$\sim$100 channels or $\sim$20~ns.  Each time window was
100~ns long.}
\label{waveform}
\end{figure}

The output pulse was sent to the VXI electronics, where the
waveform was digitized (Fig. \ref{fig:VXI}).  Different delays were added to the LED trigger to center the captured waveform in
the time windows (Fig.~\ref{waveform}). The pulses were digitized over 512 channels at
a rate of 5~giga-samples/sec using the Tektronix oscilloscope
inside the VXI crate. 

\subsection{Global testing: procedures and data handling}

The testing data were acquired in two modes.  First, the dark rate
rates were collected by recording the noise rates measured at a range
of operating voltages, with no light source. Second, the PMTs were
illuminated by an LED light source whose wavelength was 450~nm (corresponding approximately to the \v{C}erenkov frequency of the oil). The source flashed at a rate of 1~kHz; the width of each pulse was 1~ns.  PMT responses to 600-1000 LED flashes were used to determine
the operating voltage for the required gain.  This procedure also
allowed both the charge and time resolutions, and pre- and
post-pulsing anomalies to be studied.

\subsection{Global testing: results}

Each PMT was tested to determine its operating voltage, as well as its gain, dark rate, charge resolution, timing resolution, and
double pulsing rate.  The results of the later five measurements are reported below.
%Table~\ref{tab:calib} reports the spread
%in measurements, as obtained using calibration PMTs which were tested 
%multiple times.

%\begin{table}[t]
%\caption{Errors on PMT figures of merit as determined from calibration data}
%\begin{center}
%\begin{tabular}{|c|c|}
%\hline \hline
%gain & 0.19  \\ \hline
%charge resolution & 0.34 \\\hline
%timing resolution & 0.13 \\\hline
%double-pulsing percentage & 0.60 \\\hline\hline
%\end{tabular} 
%\end{center}
%\label{tab:calib}
%\end{table}

\begin{figure}[t]
  \centering
\includegraphics[width=4.in]{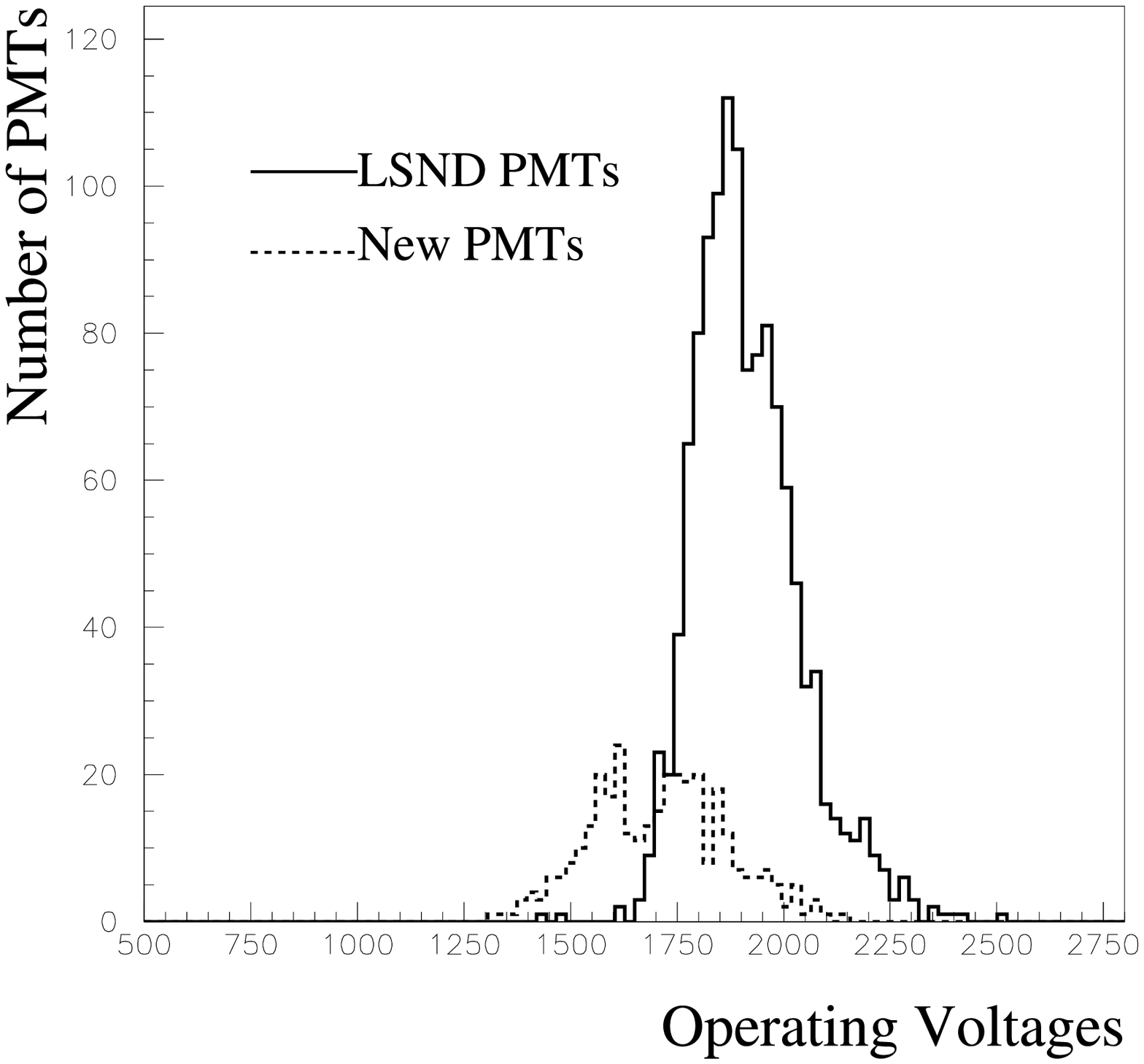}
\caption{Distribution of operating voltages for PMTs in the detector
  (from ref.~\cite{Fleming:2002mv}).}
\label{fig:opvolts}
\end{figure}

PMT gain corresponds to the mean number of electrons produced by the
phototube in response to one PE.  This was determined using the
equation
\[
gain = \left( \frac{Q_{tot}}{N} \right) \left( \frac{1-e^{- \mu}}{\mu} \right).
\label{eq:rgn}
\]
where $Q_{tot}$ is the total charge from the PMT for all pulses with response past threshold, $N$ is the number of
responses past threshold, and $\mu$ is the light level. 
  Total charge was computed by summing up
charge in the main PMT pulse for all responses with in-time pulses
that pass threshold, excluding double pulses. 
The second factor is the reciprocal of the average number of PEs seen by a PMT for a
given light level, $\mu$, (excluding zero responses) predicted from
Poisson statistics. 
 The voltage was
selected in order to obtain a gain of $1.6 \times 10^7$ electrons per
${\rm PE}$.  The distribution of operating voltages at which the PMTs
run in the detector is shown in Fig.~\ref{fig:opvolts}.

Dark rate is defined as the number of pulses larger than 3~mV in one
second.  The experiment required that the tubes operate with a
dark rate below 8~kHz in the main tank and below 4~kHz in the veto.
Dark rates were measured over a range of voltages, from several hundred
volts below suggested operating voltage up to operating voltage and
above. Plots of dark rate versus voltage, or plateau plots, serve as a measure of PMT quality.  The PMTs are operated on the plateau, where
dark rate does not change significantly as the operating voltage
increases.  The R1408 PMTs were found to have poorly defined
plateaus, but they were considered acceptable if the dark rates had a
steady, shallow rise when the operating voltage was increased.

\begin{figure}[t]
\centering
\includegraphics[width=4.in, bb= 40 150 530 625]{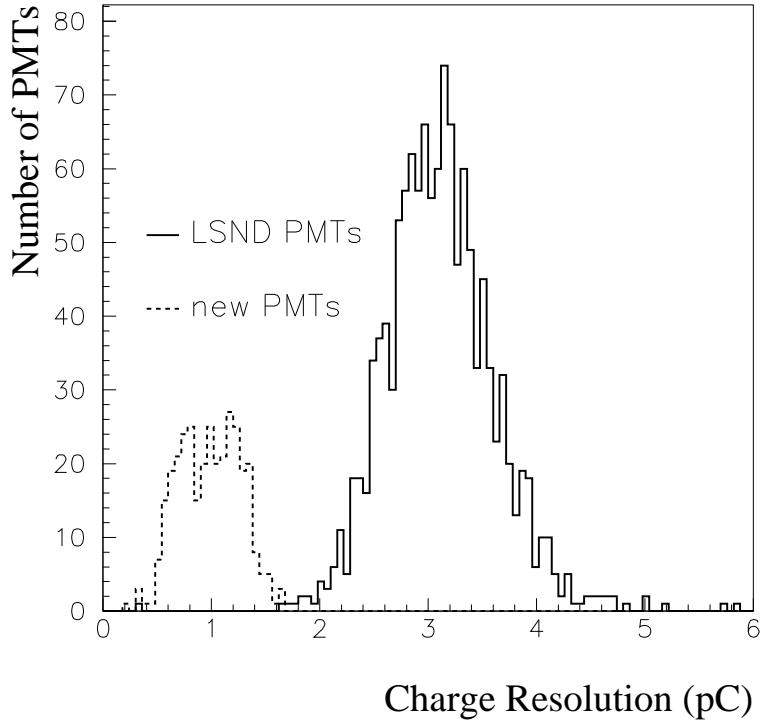}
\caption{Distribution of PMT charge resolutions (from ref.~\cite{Fleming:2002mv}).}
\label{fig:chargedist2}
\end{figure}

The charge resolution, $\sigma_q$ is determined from the width of the
peak in the PMT response corresponding to one PE.  The distribution of charge resolutions for all PMTs is
shown in Fig.~\ref{fig:chargedist2}.  

\begin{figure}[t]
\centering
\includegraphics[bb=43 155 525 618, width=4.in]{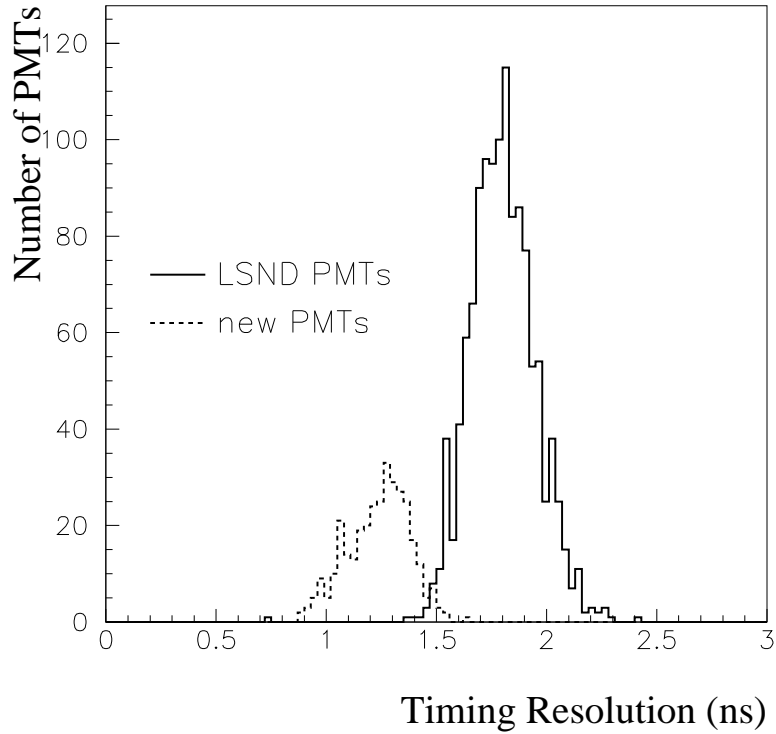}
\caption{Distribution of PMT time resolutions.}
\label{fig:timedist2}
\end{figure}

The time jitter in the
measurements is dominated by the timing resolution of the PMTs; the
timing resolution of the oscilloscope was found to be negligible. The
distribution of timing resolutions for all PMTs is shown in
Fig.~\ref{fig:timedist2}.

\begin{figure}[t]
\centering
\includegraphics[width=4.2in, bb= 37 25 519 483]{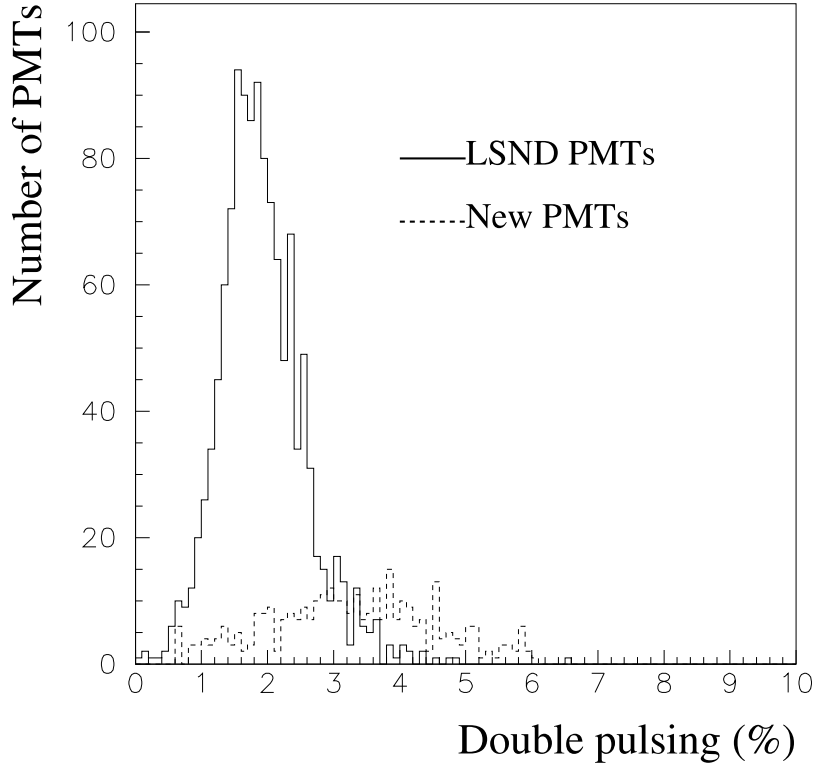}
\caption{Distribution of PMT double-pulsing rates.}
\label{fig:doubleplot}
\end{figure}

Pre- and post-pulsing, where the main PMT pulse is preceded or
followed by additional small pulses, is a known effect in the types of
PMTs used on MiniBooNE \cite{Fleming:2002mv}.  Pre-pulsing was observed in only four R1408 tubes, which were rejected.  Post-pulsing occurs in two different time
intervals: between 8-60~ns after the main pulse, and 100~ns-16~$\mu$s
after the main pulse.  MiniBooNE is concerned about the first case
(``early post-pulsing''), because data in the detector are recorded in
100~ns intervals, making it unlikely that the second case will
contaminate events.  Early post-pulsing can begin when an electron
accelerated to the first dynode starts a typical cascade, and causes
another electron to be ejected from the first dynode.  This second
electron can move around the inside of the PMT dome before settling
back to the first dynode and initiating a second cascade, which
becomes the post-pulse.  
Hamamatsu reports that R5912 and R1408 PMTs are
expected to have early post-pulses in $3\%$ of the responses from every tube. MiniBooNE
observed higher rates.  PMTs were required to have a double-pulsing
rate of $<6\%$ for the R5912 PMTs and $3\%$ for the R1408 PMTs at the time of installation. The
distribution of PMT double-pulsing rates for all PMTs is shown in
Fig.~\ref{fig:doubleplot}.

\section{Angular Testing}
\label{sec:angular testing}
Extensive specialized testing was done on seven phototubes to explore
the behavior of other important parameters, especially the dependence
of the PMT response on the angle of the incoming light.  The data from R1408 and R5912 tubes were kept separate until the final analyses. The main objective of the angular tests was to find how the response of the PMTs depended on the angle of incident light. There were three additional tests. One tested a few tubes at various voltages. Another rotated the tubes about two orthogonal axes arbitrarily labelled ``yaw" and ``pitch" to test the assumption that the tubes are rotationally symmetric. The last tested tubes in air, without the oil present in the other tests.

\subsection{Angular tests: apparatus}

The apparatus used for these tests is designed to replicate as closely as possible the characteristics of  the MiniBooNE detector, on a much smaller and more manageable scale. The
central part of the apparatus is a 40 gallon, stainless steel tank,
 painted on the inside with the paint used in the detector tank: a primer and black top coat. The primer is
Hydralon P water-based epoxy primer (Sherwin Williams
E72AC500/V66VC503 Hydralon P water-based epoxy primer), and the black
topcoat is a flat black moisture cure aliphatic polyurethane chemical
agent resistive coating (F93B102 Flat black moisture cure
polyurethane). The tank can be filled with MiniBooNE oil when tests
require it.  It houses a single PMT, which can be rotated along the polar or azimuthal angle of the tube, using two precision rotary tables, one for pitch and
one for yaw with respect to the table. The tubes were fixed to these tables with an arbitrary rotation, so ``pitch" and ``yaw" refer only to the mechanical apparatus and not any characteristic of the tubes.  These tables have externally-coupled  mechanical controls which allow the phototube position angle to be changed without opening the tank and exposing the tube to light. A 10.25 inch diameter window at the front of the tank allows the tests to simulate the light in MiniBooNE events by propagating the light several meters before it strikes the
phototube.  Because of this window, the entire
 testing room must be kept dark throughout each test.  

%dude there's a story here 
 
The light source is a PicoQuant PLS 450, a  sub-nanosecond pulsed LED.  Its center wavelength is 460~nm, with 50~$\mu$W average power at 40~MHz,
approximately 30~nm spectral width, and typically 800~ps pulse width.
It was used in 10 second sample periods at about $10^6$~ Hz. The light was filtered to produce a low-intensity beam ($<$1 photoelectron per pulse). The light was directed through an optical cable to face the tank from a distance of approximately 3~meters. The light from this source makes a cone with half-angle about 23$^{\circ}$, with greatest intensity along the centerline.  The 3~m distance is great enough that one can assume that the entire face of the tube is uniformly  illuminated with light in parallel rays. 
 
%The phototube can be mounted so that it pivots around the center of
%curvature of the face (thus the center of the light beam enters normal
%to the face but at a varying location) or so that it rotates around a
%point at the center of the face of the tube (thus the light enters at
%a varying angle of incidence but constant center point). For the angle
%dependence tests, the first mounting was used, and the subsequent
%modeling of the tube response takes this into account.

%%Since this isn't what we ended up doing... we should cover our tracks
%Mini-MiniBooNE has the capability to perform
%a point-by-point mapping of possible structural effects in individual
%PMTs.  This is accomplished by pulsing a collimated, low level ($<$10
%photoelectrons) light source at at close range (within the
%Mini-MiniBooNE tank) at varying angles and positions of incidence at
%points across the PMT face. This map could then be unified into a
%description of the response from the whole tube. The full MiniBooNE
%Monte Carlo accounts for the PMT face geometry, and so this type of
%description is already internalized within the simulation.  To avoid
%redundancy and simplify testing, tests were performed by simulating an
%actual event with light striking the entire face of the PMT from
%several meters away. The results could then be normalized by dividing
%out the Monte Carlo geometry simulation.

\begin{figure}[htbp]
\centering
\includegraphics[bb= 93 532 535 708, width=.7\textwidth]{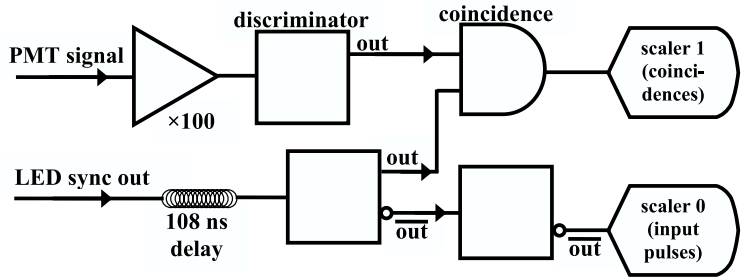}
\caption{Testing tank data acquisition system block diagram.~\cite{Justin IEEE} }
\label{mmboone daq schematic}
\end{figure}

The signals from the tube in the testing tank can be read out via an oscilloscope/GPIB card, or via the MiniBooNE
data acquisition system. For the angular tests, a specialized data aquisition system was used (Fig. \ref{mmboone daq schematic}). The  signal from the tube in the tank is
amplified by a factor of 100, sent to a discriminator with 30 mV
threshold, and finally put into coincidence with the delayed sync out
signal from the LED pulser. The number of coincidences and the number
of input light pulses each go to a scaler; they are then recorded and
used to calculate the relative efficiency. 

%The decision was made to shield the tank magnetically after concerns
%were raised about magnetic fields influencing the observed PMT
%response. Rotating the entire structure (while the alignment of PMT,
%collimator, and internal light source remained constant) showed an
%effect from ambient (i.e., the Earth's) magnetic fields of the same
%order of magnitude as the observed structural effects. However, at the
%level of these effects, the magnetic component could not totally
%account for the observed response, so magnetic shielding was used in
%all subsequent testing.

The tank is magnetically shielded with a $\mu$-metal box. This was done because ambient magnetic fields (e.g., the Earth's) can influence the relative efficiency of the tubes, and thus rotation within these fields may change the shape of the relative efficiency results.  To see if ambient magnetic fields could have a significant effect, the tank was rotated (while holding the configuration of the tube and internal light source constant). The resulting effect was smaller than (but on the same order as) the angular dependence. When the tank was shielded and the same tests were performed, this dependence on orientation within ambient fields was not seen.

%The tank is magnetically shielded with a $\mu$-metal box.
%The tank is shielded from the possible influence of ambient magnetic fields (e.g., the Earth's) on the observed PMT angular response, since
%the measurement of that response involved rotating the tube itself.
%This was done after tests showed that rotating the tank (while holding the configuration of the tube and internal light source constant)
%showed a smaller (but on the same order of magnitude) change as the
%previously observed tube rotation angular effect.  It was not the
%case, however, that these external magnetic effects could entirely
%account for the observed response. To obviate potential complications
%from competing effects, therefore, the tank was magnetically shielded by means of a $\mu$-metal box.

\subsection{Angular tests: procedures}
The apparatus detailed above  was used to study relative efficiency ($E_{rel}$) as
a function of angle. Using rapid, short-duration pulses of the LED,
the $E_{rel}$ of the tube was measured at a given position by taking the
ratio of the number of responses which coincided with LED pulses to
the total number of pulses, with about $10^7$ pulses per datum.
\[
\left(
\frac{ \mbox{coincidences out}}{\mbox{pulser pulses in}}
\right)
 = \mbox{relative efficiency ($E_{rel}$)}\]

 $E_{rel}$ values were measured at $5^{\circ}$ increments through the
 entire $360\,^{\circ}$ of response, with two repetitions at each
 point and two rotations of the tube (from $\theta =-180^{\circ}$ to
 $\theta =+180^{\circ}$ and back again) for a total of four
 measurements at each tube position. Data were recorded directly into
 a spreadsheet, providing an ongoing real-time check.

 This rotational test was performed on four R1408 and three
 R5912 tubes.  Most of these tubes had been previously
 tested with the global testing (see Section \ref{sec:global
   testing}), so the expected dark rate was known, and known to be
 comparable to that for the other tubes. The global testing results
 also specified tube operating voltages. It was necessary to dark
 adapt each tube for at least 12 hours to obtain a dark rate
 comparable to that of the tubes in the detector, which have not been
 exposed to bright light for over three years.

%five for silver, six for gold, seven for the secret never to be told...

%deleted: repetitious
% Since each of the tubes which was used in specialty testing had
% previously been generally tested, we knew what ballpark dark rate to
% expect, and we knew that this rate was similar throughout the
% detector.

\label{pgh: roll doesn't matter}
The phototubes are essentially rotationally symmetric about their central vertical
axis, with variation only in the dynode structure below the main dome.
The R1408 tubes have ``venetian blind'' style dynodes; the R5912
tubes have ``box and line" style dynodes. No significant differences
were observed when the dynodes at the starting position were oriented
horizontally, vertically, or at $45^{\circ}$.  Additionally, a test
was performed on one tube varying the yaw angle (similar to the pitch
measurements performed on all tubes); again no significant differences
were observed. The equivalence of pitch and yaw implies that that the
roll orientation of the tube does not affect the results. This
simplified the measurements greatly.
 
%(this is high on the list of things to cut, but it's here for discussion:)

%A test was performed to check the assumptions that the distance of the
%light source from the phototube had very little influence on the
%response, beyond the expected $\frac{1}{r^2}$ scaling factor, and that
%the light source was centered on the tank window. The optical fiber from
%the light source was moved 25 cm closer to the tank and
%12 cm down. While the light source was very precise in
%time, it had an unpredictable spatial distribution which was
%consistent only when not physically moved, as it was in these tests
%(and these tests only).

Another test was performed by varying the operating voltages by 50~V
above and below the nominal voltages found by previous MiniBooNE
tests. The gain variation with voltage, which determines the
$\max E_{rel}$, was already known, so this test checked for variation in
the shape of the $E_{rel}$ curve more than the expected variations in
amplitude. None of significance was observed.

%skip this
% that our data for a
%particular run looked like it matched previous runs. This proved
%extremely useful---for example, when the data from one tube seemed to
%be peaking at about half the normal angle and with half the normal
%peak height, we discovered that the phototube had become disengaged
%from the rotating apparatus, and was bobbing freely in the oil tank.
%If we had continued testing, the apparatus and phototube would have
%been significantly stressed and possibly broken.

\vspace{0.1in}

\subsection{Angular tests: results}

%order of testing:
% july 12 15s3
% july 19 20s1 air
% july 20 20s1 oil
% july 21 20s1 change V
% july 25 SA2272 oil with change V
% july 27 20n16 oil
% july 28 19s3 oil
% augus 3 SA2753 oil
% augus 4 SA2762 oil
% augus 6 SA2761 oil

\begin{figure}
\vspace{5mm}
\centering
\includegraphics[width=10.cm, clip=true]{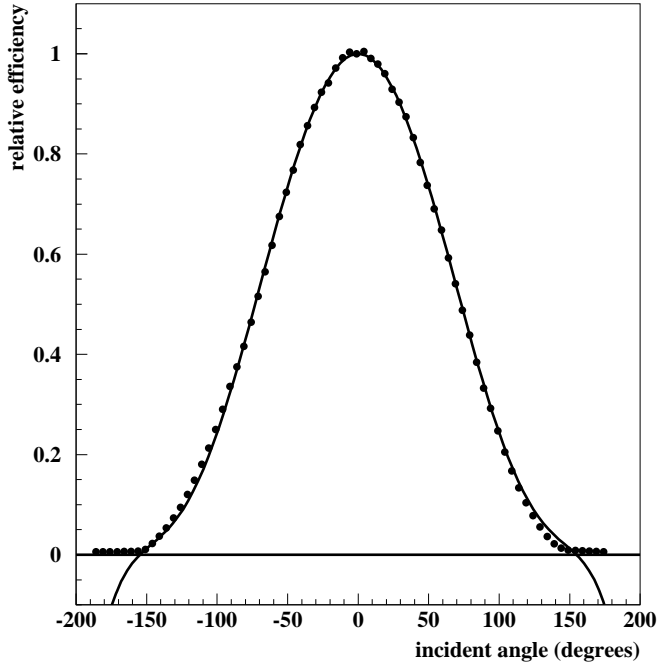}
\caption{A typical data set from an R1408 tube (19s3) with its
  symmetric polynomial fit. This fit uses the horizontal shift and maximum relative efficiency found from the initial quadratic fit.}
  \label{pmtgraphold}
\end{figure}

Plots of relative efficiency as a function of angle were made for each
set of measurements; these are roughly bell-shaped, as one would
expect from the shape of the tubes.
Typical results from an R1408 and an
R5912 (with the fits as described below) are included in Figures
\ref{pmtgraphold} and \ref{pmtgraphnew}.
%and \ref{pmtgraphnew}. 
 These plots are similar from tube to tube
(see Figure \ref{pmtgraphall}).  
The parameters for the fits, as well as the average which will be used
in the MiniBooNE Monte Carlo simulation, are included in Table \ref{table:old parameters}.
  % and \ref{table:new parameters}. 
The results obtained were compared to expectations from  the shape
of the PMTs and results obtained by the SNO~\cite{SNO}
experiment.

\subsubsection{Angular tests: results and polynomial fits}

\begin{figure}
\vspace{5mm}
\centering
\includegraphics[width=10.cm, clip=true]{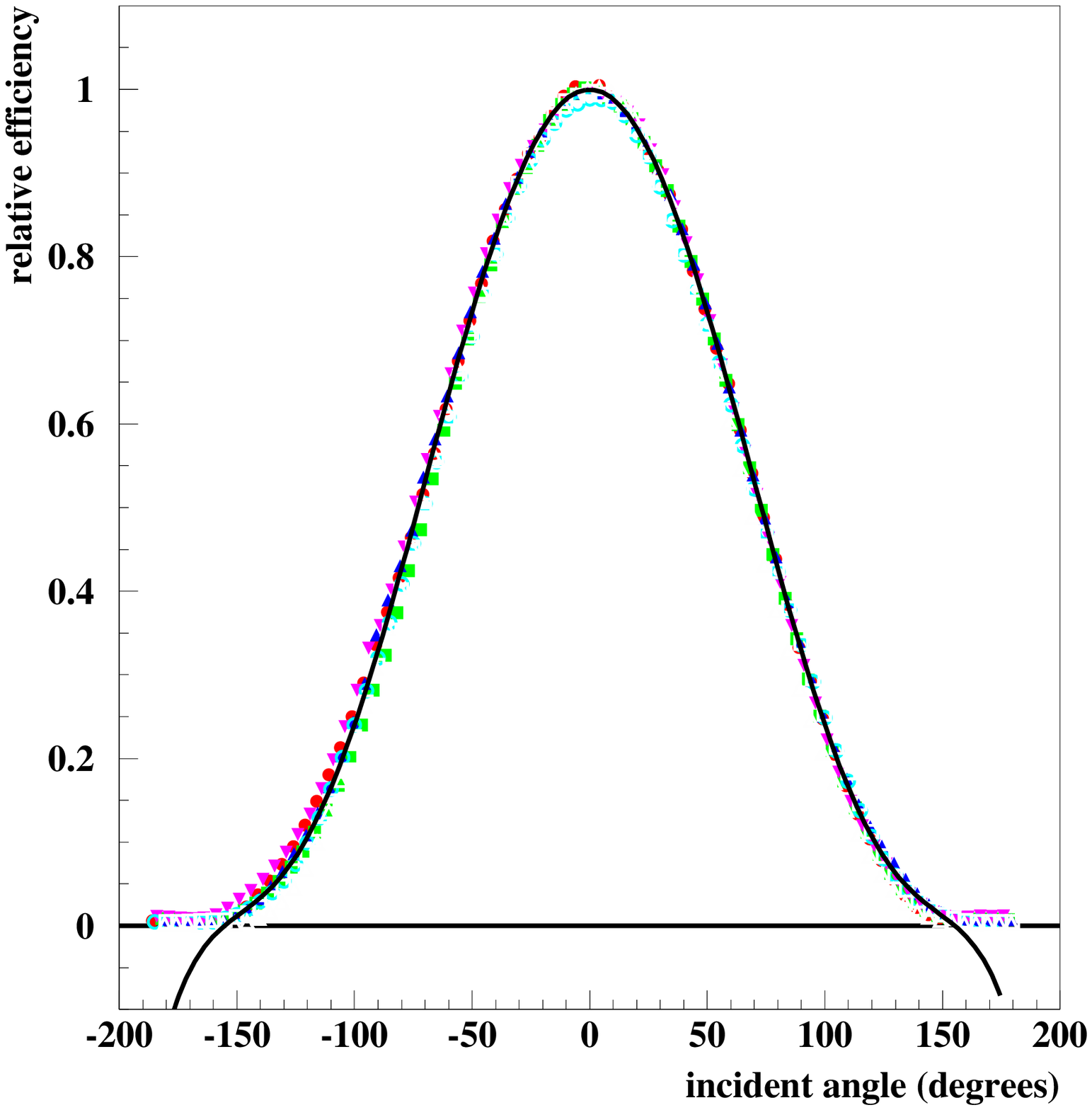}
\caption{Composite relative efficiency data. The data in each
  series were fitted to a polynomial, and normalized to $\max E_{rel}= 1$   at $\theta=0^{\circ}$. 
%Each data series is labelled by PMT serial number: the R1408 PMTs are numbers 15s3, 20s1, 20n16, and 19s3; R5912 PMTs are   SA2272, SA2753, and SA2761. 
The average fit is shown.}
%Note that for positive angles, (i.e., the tube is facing towards the surface of the oil in the Mini-MiniBooNE tank) the new tubes have consistently higher $RE$ than the old tubes. One speculative explanation of this could be that this is related to internal reflection from the surface of the oil, and since the new PMTs have higher gain than the old ones, they are more sensitive to this difference. This would predict a larger error in positive angles, but it would suggest that the data should be higher than the fit, and ours is typically lower.
\label{pmtgraphall}
\end{figure}

Because the PMT rotator mount inside the tank does not precisely
specify the initial angle, $\max E_{rel}$ is found for each curve; the
angle corresponding to $\max E_{rel}$ is assumed to correspond to
$\theta=0$. Each plot's horizontal axis is shifted accordingly before the plots are compared.  Each y-axis is scaled to
$\max E_{rel}=1$. An initial quadratic fit is performed on the central points of each plot to determine the  horizontal shift. A 6-degree  polynomial fit is then performed on each plot in the region $\theta=-150^{\circ}$ to $\theta=150^{\circ}$. Outside of this region, statistics were significantly lower, making the data less comparable from tube to tube. The fit
polynomial is forced to be symmetric by employing only even powers of the incident angle. The polynomial coefficients
are varied to minimize the $\chi^2$. Finally, before comparing the
functions, each fit is normalized by dividing by its constant term
(which equals $\max E_{rel}$), forcing $\max E_{rel}=1$. The results of the fits are shown in Table \ref{table:old parameters} and superimposed on Figures \ref{pmtgraphall} and \ref{pmtgraphairsno}.

\begin{figure}
\vspace{5mm}
\centering
%, clip=true
\includegraphics[width=10.cm, angle=0, clip=true]{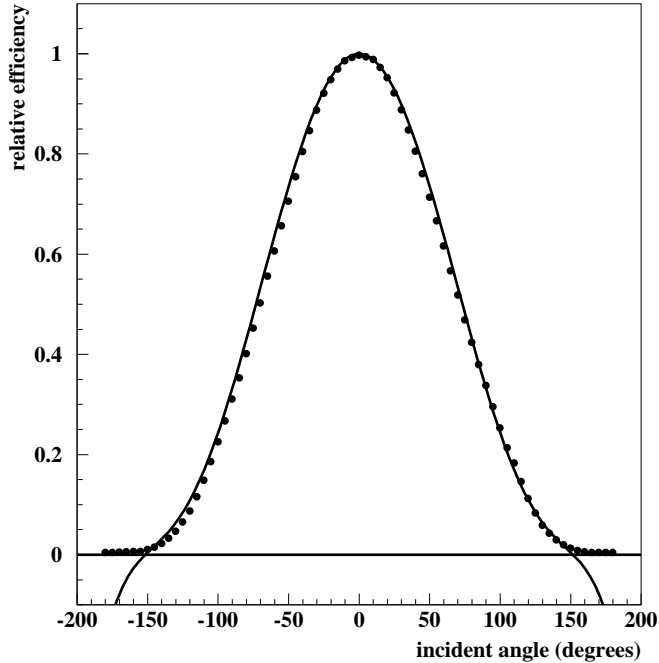}
\caption{A typical data set from an R5912 tube (SA2761) with its
  symmetric polynomial fit, analogous to Figure \ref{pmtgraphold}.}
  \label{pmtgraphnew}
\end{figure}

\begin{figure}
\vspace{5mm}
\centering
\includegraphics[width=10.cm, clip=true, angle=0]{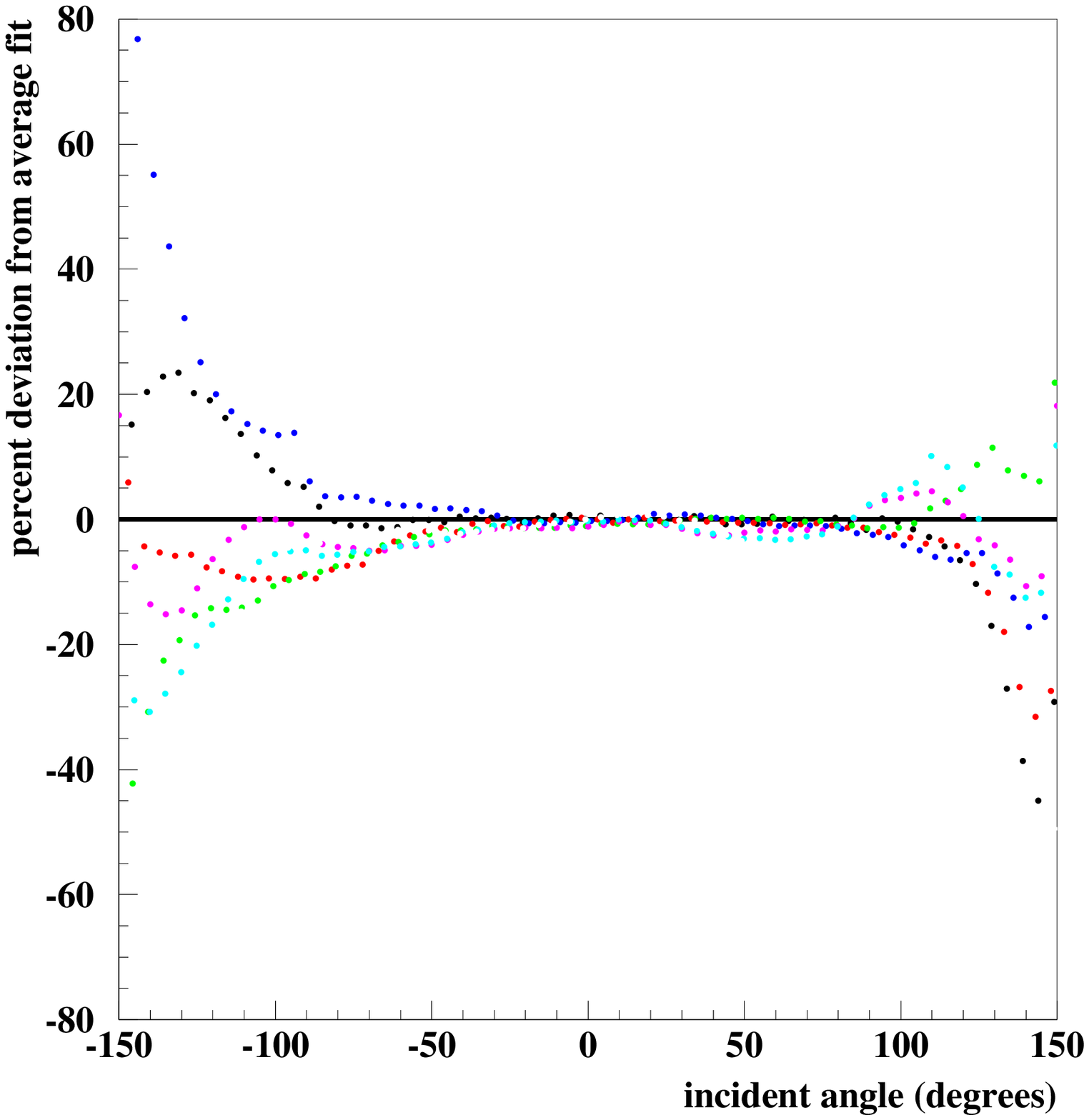}
\caption{Comparison of 
  the data for each tube to the average fit. This
  plot shows, for each tube, the percent difference between the data and
  the average fit (i.e., $100\times(E_{rel}-f(\theta))/f(\theta)$).}
  \label{pmtgraphdiff}
\end{figure}

The errors on these data are dominated by tube-to-tube variation. The
tests on each tube individually had very high statistics, typically
around $2 \times 10^7$ light pulses at each measured angle, and about 600~000 responses at central angles or 4~000 at extreme angles (where the tube was facing backwards). 
Thus, to describe the error, it is more important to look at the
relatively large differences between data sets from different tubes.
This was done by examining the parameters of the fits. 
The average of the seven fitted values of
each parameter and their standard deviation from that average are given
in Table \ref{table:old parameters}.  Figure \ref{pmtgraphall} compares each PMT to the average fit, and
deviations from the average fit are shown in Fig. \ref{pmtgraphdiff}.

\begin{table}
  \caption{%Normalized R1408 (old) tube parameters.
    Normalized parameters from all all tubes tested. The error is the standard deviation from the average of the seven tubes tested. These parameters are to 
    be used in the following function for the Monte Carlo simulation, where $\theta$ is 
    the angle of the incident light from the tube axis measured in degrees.} 
 \[E_{rel}= 1 + a_2\theta^2 + a_4\theta^4 + a_6\theta^6.\]
 \begin{center}
\begin{tabular}{|l|r|r|r|r|}
\hline \hline
tube number & $a_2$& $a_4$ & $a_6$\\
\hline
15s3 &    -1.182$\times 10^{-4}$&5.018$\times 10^{-9}$&-7.624$\times 10^{-14}$\\ 
19s3&     -1.234$\times 10^{-4}$&5.436$\times 10^{-9}$&-8.468$\times 10^{-14}$\\
20s1 & -1.162$\times 10^{-4}$&4.818$\times 10^{-9}$&-7.136$\times 10^{-14}$\\
20n16&   -1.156$\times 10^{-4}$&4.725$\times 10^{-9}$&-6.728$\times 10^{-14}$\\
SA2753&   -1.101$\times 10^{-4}$&4.191$\times 10^{-9}$&-5.539$\times 10^{-14}$\\
SA2761&   -1.177$\times 10^{-4}$&4.964$\times 10^{-9}$&-7.552$\times 10^{-14}$\\
SA2272&  -1.264$\times 10^{-4}$&5.563$\times 10^{-9}$&-8.549$\times 10^{-14}$\\
\hline
average&  -1.182$\times 10^{-4}$&4.959$\times 10^{-9}$&-7.371$\times 10^{-14}$\\
error& 5.337$\times 10^{-6}$& 4.583$\times 10^{-10}$&1.042$\times 10^{-14}$\\
\% error& 4.5 & 9.2& 14.1\\
\hline
\hline
\end{tabular}
\label{table:old parameters}
\end{center}
\end{table}

%\vspace{20pt}
%Table \ref: Pre-normalized R5912 tube parameters. These parameters are to be used in the following fit function, where $\theta$ is the pitch angle measured in degrees. The $a_{iL}$'s are to be used for $\theta <0$, and the $a_{iR}$'s are to be used for $\theta>0$.
% \[RE= a_0 
% + a_1\left(\frac{\theta-xshift}{100}\right) 
% + a_2\left(\frac{\theta- \theta_{shift}}{100}\right)^2 
% + a_3\left(\frac{\theta- \theta_{shift}}{100}\right)^3 \cdots\]

%\begin{tabular}{|l|r|r|r|r|r|r|}
%\hline
%\hline
%tube number &  $\theta_{shift}$& $a_{1L}$& $a_{2L}$&$a_{3L}$&$a_{4L}$&$a_{5L}$\\
%           &$a_0$ &$a_{1R}$&$a_{1R}$&$a_{1R}$&$a_{1R}$&$a_{1R}$\\
%\hline
%SA2753&  1.99992 &0.01593& -0.07257   & 0.00389 &0.04783 &0.01543\\
%&0.07601& -0.00008&    -0.08130&    -0.01015&    0.05159 &-0.01560\\
%\hline
%SA2761 & 1.99954 &-0.01712 &   -0.16178  &  -0.10416 &   -0.01414 &   0.00210\\
%&0.07487& -0.00063&    -0.12106  &  0.06831& 0.00063 &-0.00455\\
%\hline
%SA2762 & 1.99999& 0.02478& -0.05451 &   0.00268& 0.04052& 0.01349\\
%&0.06657 &0.00329& -0.07103 &   -0.01124   & 0.04447& -0.01288\\
%\hline
%SA2272 & 1.99994 &0.00399& -0.05249&    0.01079 &0.03845 &0.01140\\
%&0.05226 &-0.00360  & -0.06464  &  -0.00552 &   0.04679 &-0.01568\\
%\hline
%Average &1.99985& 0.00690& -0.08534 &   -0.02170  &  0.02817 &0.01060\\
%&0.06743 &-0.00026&    -0.08451   & 0.01035& 0.03587& -0.01218\\
%\hline
%\hline
%\end{tabular}

The even-polynomial fits are sufficiently
accurate and consistent for the R1408 tubes. For the R5912
tubes, the fits are less accurate on one side (see Figure
\ref{pmtgraphnew}). To describe this difference, several other fits of the R5912 tube data were attempted.  The main attempted
non-symmetric fits used a different 5-degree polynomial for angles
above or below zero, requiring only that they meet at $\theta=0$. The same horizontal shift and $\max E_{rel}$ were used for this fit as for the initial, even polynomial fit. However, it was found that the
variation between individual tubes was larger than the average
difference between R5912 tubes and R1408 tubes, so the even polynomial fit was used for all tubes. 
%typical data set for an R1408 and an R5912 with their fits are shown
%in Figures \ref{pmtgraphold}, \ref{pmtgraphnewsym}, and \ref{pmtgraphnew}.

\subsection{Angular tests: comparison to predictions and to SNO data}

\begin{figure}[htbp bbllx=75bp,bblly=57bp,bburx=534bp,bbury=733bp]
\centering
\includegraphics[width=10.cm, clip=true]{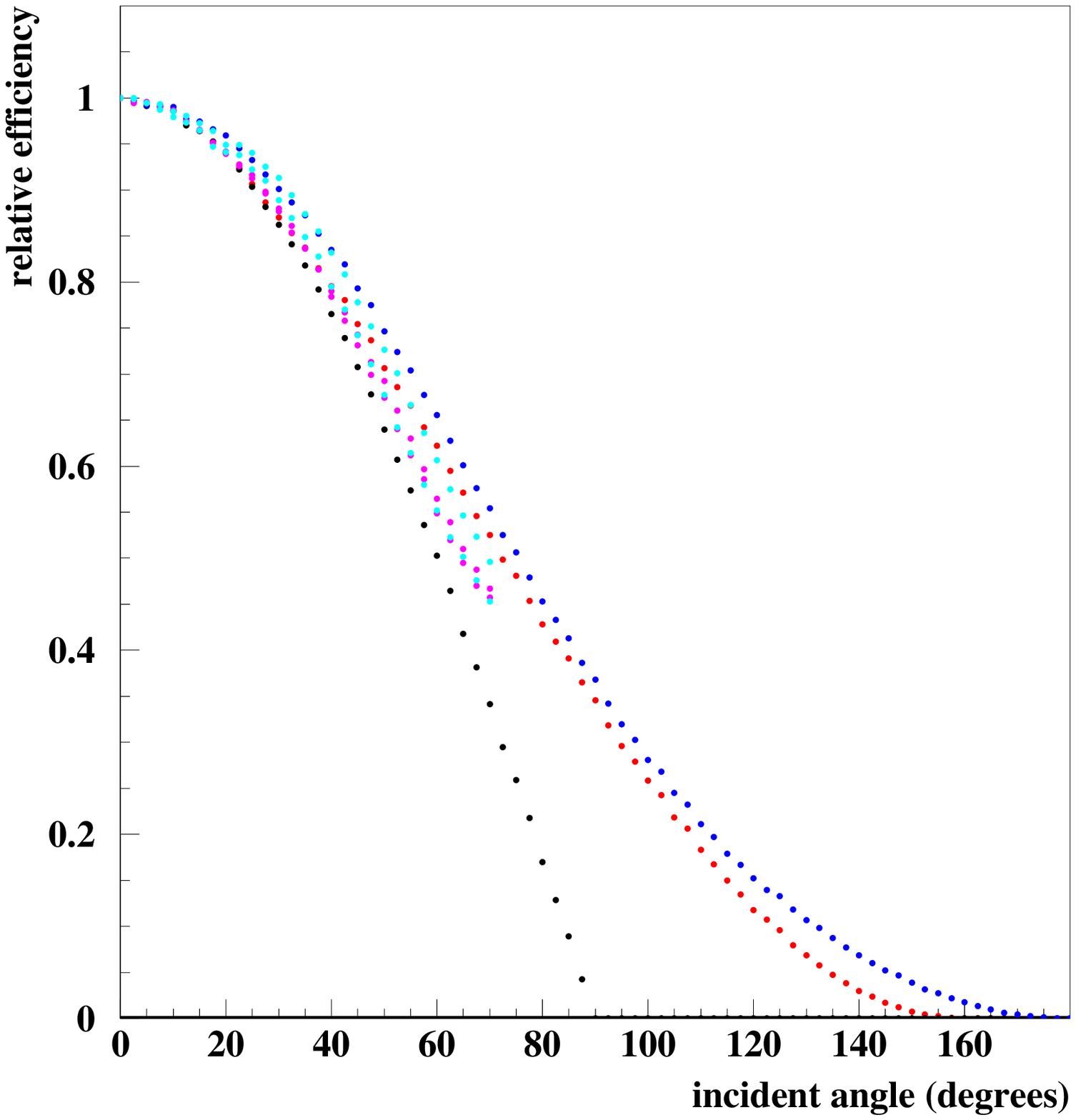}
\caption{Relative efficiency predicted by assuming that the PMT is a flat disk (black), a  hemisphere (red), or ovoid shaped as in the Hamamatsu
  technical drawing (figure~\ref{hamamatsu_drawing_dynode}) (blue), and the  results found by SNO from testing two PMTs each in air (purple) and water (cyan).}
\label{pmtpredictions}
\end{figure}

MiniBooNE's relative efficiency angular data were compared to several
geometrical models, and to data from previous tests from the
Sudbury Neutrino Observatory (SNO) experiment~\cite{SNO}. The three
geometrical models assumed the photocathode was a flat disk, a hemisphere, and the shape from the Hamamatsu technical
specifications. The data from SNO were taken from tests on two R1408 PMTs in air and water, for a total of four tests.  The PMTs in
their detector are all Hamamatsu R1408 PMTs (the same as the MiniBooNE
R1408 tubes), making comparisons with them particularly useful. These
seven results are shown together in Figure \ref{pmtpredictions}. Note
that all of the SNO results lie between the hemisphere and the ovoid
shape predictions on the top and the flat disk prediction on the
bottom. SNO used water
instead of oil because the SNO detector contains water.

\begin{figure}
\vspace{5mm}
\centering
\includegraphics[width=10.cm, clip=true]{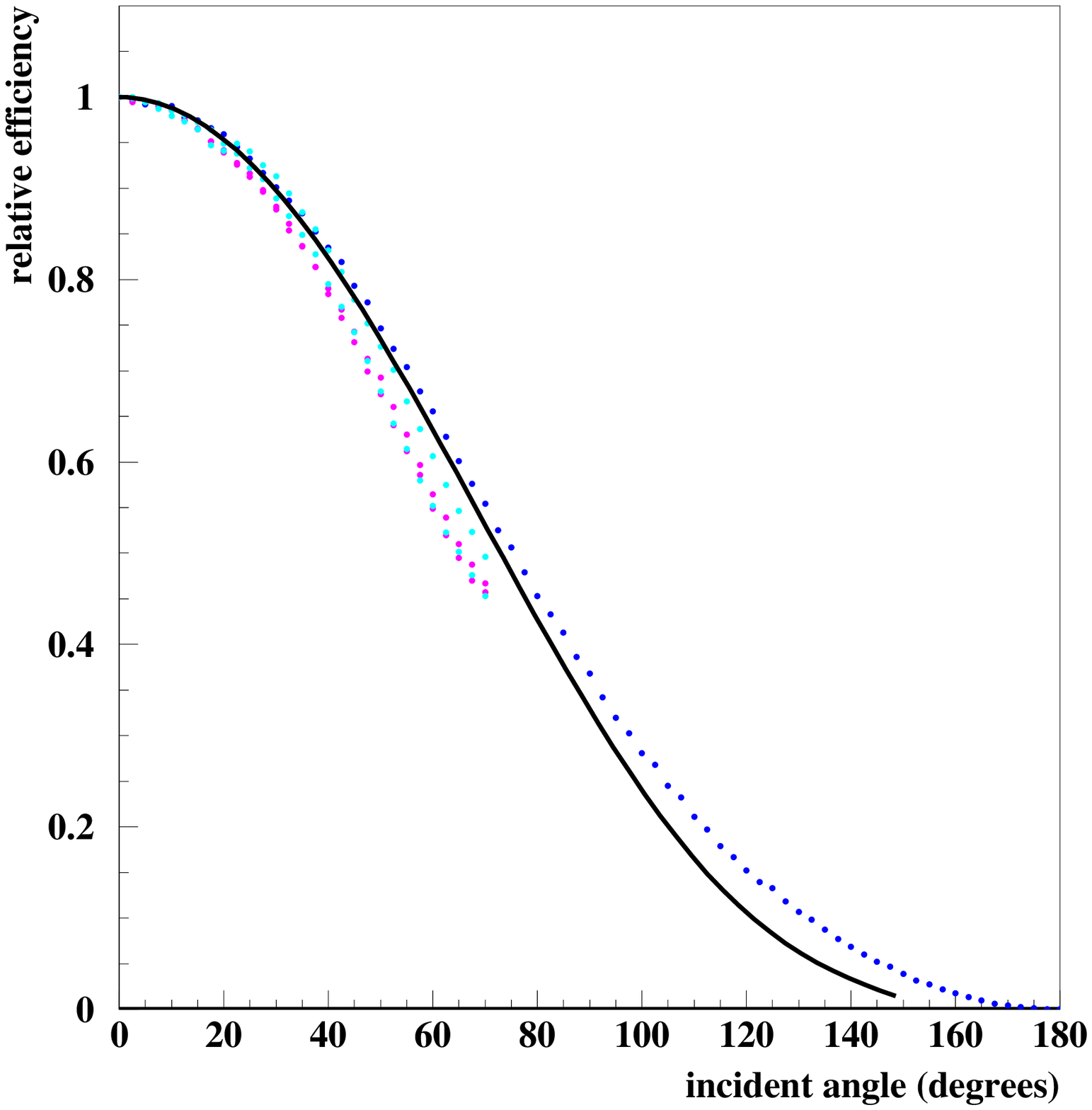}
\caption{Comparison of the average MiniBooNE $E_{rel}$ function (black curve) with the SNO results from testing two PMTs each in air (purple) and water (cyan) and the geometrical
  $E_{rel}$ function based on the Hamamatsu drawing (blue points).}
  \label{pmtgrapholdrealsno}
\end{figure} 

\begin{figure}
\vspace{5mm}
\centering
\includegraphics[width=10.cm, clip=true]{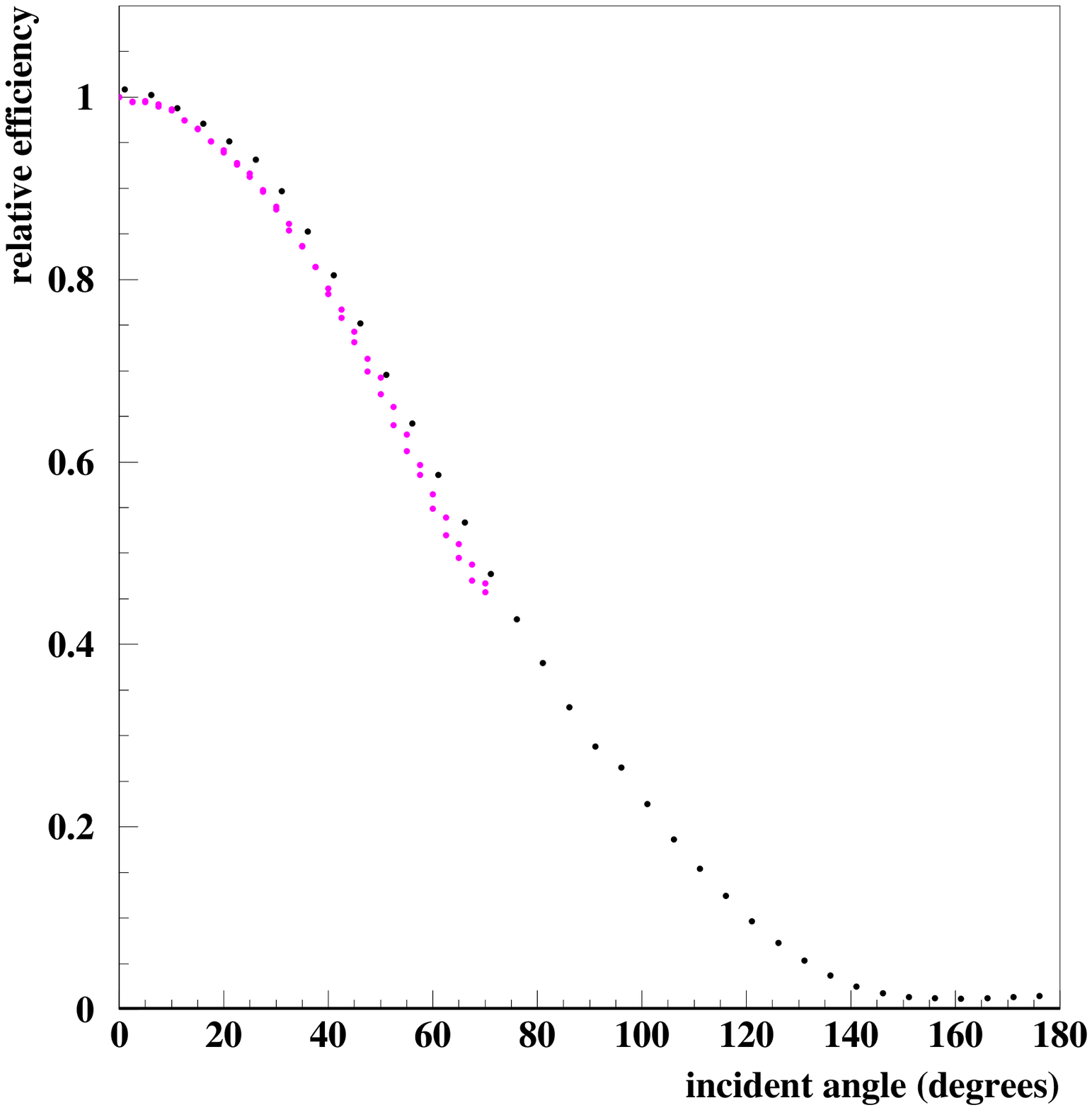}
\caption{Comparison of MiniBooNE data from the tube tested in air (black) to SNO  test data taken in air (purple). } \label{pmtgraphairsno}
\end{figure}

While developing the testing procedure, one tube was tested without filling the testing tank with oil. The results from this test, normalized to
$\max E_{rel}=1$, were very similar to the air tests done by SNO. This
comparison is shown in Figure \ref{pmtgraphairsno}. 

Comparisons among the angular response measurements, SNO data, and
geometrical models reveal overall similarities. The
present data are also observed to be closer to the geometrical model from the Hamamatsu ovoid shape description than to the results
obtained in either air or water by SNO. This may be the case because
these data were taken in oil, where the index of
refraction is much closer to that of glass than either air or water,
resulting in less reflection at the  oil-glass interface.

\section{Implosion Risk Studies}
\label{sec:implosion risk studies}

On November 12, 2001, a PMT implosion in the Super
Kamiokande(Super-K)\cite{super-k}\cite{super-k news} detector started
a chain reaction that resulted in the loss of several thousand tubes.
At the time, the MiniBooNE detector was not yet filled with oil.
Assessment of the possibility of a similar accident occurring at
MiniBooNE was immediately undertaken.

\subsection{The Super-K detector design}

%1atm= 101.325kPa
The Super-Kamiokande detector is cylindrical, 36 m tall and 34 m in
diameter, and filled with water. Hamamatsu 20-inch PMTs line the
interior on a support structure. They are spaced 0.8~m center to
center to give 40\% photocathode coverage in the tank. These tubes are
rated for a pressure of 608 kPa (6 atm). When full, the static
pressure at the bottom of the detector is 208 kPa (2.05 atm).

At the time of the accident the tank was being refilled following some
maintenance work, and was approximately $2/3$ filled with
water.  The failure began with the implosion of a single tube on the
detector floor; the reason for this failure is not known. This set off
a chain reaction which ultimately destroyed most of the tubes more
than 5~m below the water line. All tubes within 5~m of the surface
survived.

\subsection{Comparison of Super-K and MiniBooNE implosion scenarios}

MiniBooNE has a significantly smaller (12~m diameter) spherical
detector, using 8 inch Hamamatsu PMTs rated for 709 kPa (7 atm)
pressure. The tubes are spaced 0.55 m center to center. The detector
is filled with oil, which is a factor of 0.9 less dense than water.
The static pressure at the bottom of the MiniBooNE tank is 208 kPa
(4.5 atm).  Each of these factors suggests that the MiniBooNE
detector should be more resistant to a chain reaction of implosions
than was the Super-K detector.

A computational model to calculate the shock wave pressure resulting from tube implosion was developed, based upon the stored energy
in an evacuated tube at a given depth, assuming the pulse length is
the tube radius. It should be noted that the energy stored in an 8 inch
MiniBooNE tube is over an order of magnitude less than in a 20 inch
Super-K tube.

The pressure on a neighboring tube when a tube at the bottom of the
MiniBooNE detector implodes was compared to the pressure resulting
from a Super-K implosion at a depth of 5 m, the depth above which all
Super-K tubes survived. It was found the pressure in MiniBooNE was
0.45 times the Super-K pressure at survival depth, thus indicating a
safety factor of two for MiniBooNE if the tubes were equally strong.
In fact, the 8 inch tubes are rated for a higher pressure than the 20
inch tubes. MiniBooNE is therefore confident that the experiment is
operating with a safety factor of greater than two, and that an
accidental implosion of one of MiniBooNE's tubes will not result in a
chain reaction.

A more rigorous hydrodynamic calculation was done using numerical
techniques. From these calculations a somewhat greater safety factor
of 2.6 can be inferred.

\subsection{Tests performed at Super-K and SNO}

After the Super-K accident, several tests were done at Super-K using
the 20 inch tubes. Personal communication from several Super-K
collaborators indicate that at a depth of 30 m of water chain reaction
implosions occurred, but at 15 m there were no chain reactions. Since
the pressure at 15 m is significantly greater than the maximum
pressure in MiniBooNE this supports the conclusion that MiniBooNE
tubes are not at risk for this sort of accident.

In 1990, the SNO collaboration also performed a series of tests of 8
inch PMTs in a water-filled pressure vessel. No chain reactions
occurred at a pressure of 608 kPa (6 atm). This test also supports the
previous conclusion.

\section{Conclusion}

This paper has reported on tests of the R1408 and R5912
photomultiplier tubes used in the MiniBooNE experiment.  We have
briefly described the global testing of all tubes, described in detail
in reference \cite{Fleming:2002mv}.  The main results reported in this
paper pertain to the phototube angular tests, and
to the calculation of implosion risk for MiniBooNE.   The angular response of the tubes determined a sixth
order polynomial function describing the angular response.  Implosion
risk calculations for the tubes in the MiniBooNE detector were
described.  It was found that the MiniBooNE detector has a large
safety factor.

\section{Acknowledgments}
The authors would like to thank the following people for their
valuable contributions: Jesse Guerra, Andy Lathrop, Zhijing Tang, Ryan Patterson, T.
Neil Thompson, Sally Koutsoliotas, and Christi Bohmbach. Thanks are due
also to the following people for helpful discussions and editing:
Bruce Brown, David Finley, Bill Louis, Peter Meyers, Mike Shaevitz, Ray
Stefanski, and Morgan Wascko.

This material is based in part upon work supported by the National
Science Foundation under Grants No.  NSF
PHY-0139464 %(Umich REU (continuing since 2002 to at least 2005)),
and NSF PHY 00-98826 % (other REU for Christi Bohmbach)
and the U.S. Department of Energy No.
DE-FG02-91ER40671 %princeton's general grant(from Peter Meyers)
.

%"NSF support also must be orally acknowledged during all news media interviews, including popular media such as radio, television and news magazines." interesting...

%The low light studies were performed by Christi Bohmbach, with 
%advisor Darrel Smith (Embry Riddle).
%T. Neil Thompson for contributing to the base design.
%S. Koutsoliotas for contributing to the global testing.
%The rigorous hydrodynamic calculations for the implosion-risk study were done by Zhijing Tang (FNAL engineer).

 \appendix

% \section{}
% \label{}

\section{Hamamatsu Design Specifications}

\begin{figure}[ht]
\includegraphics[width=5.45in, bb= 0 0 596 792]{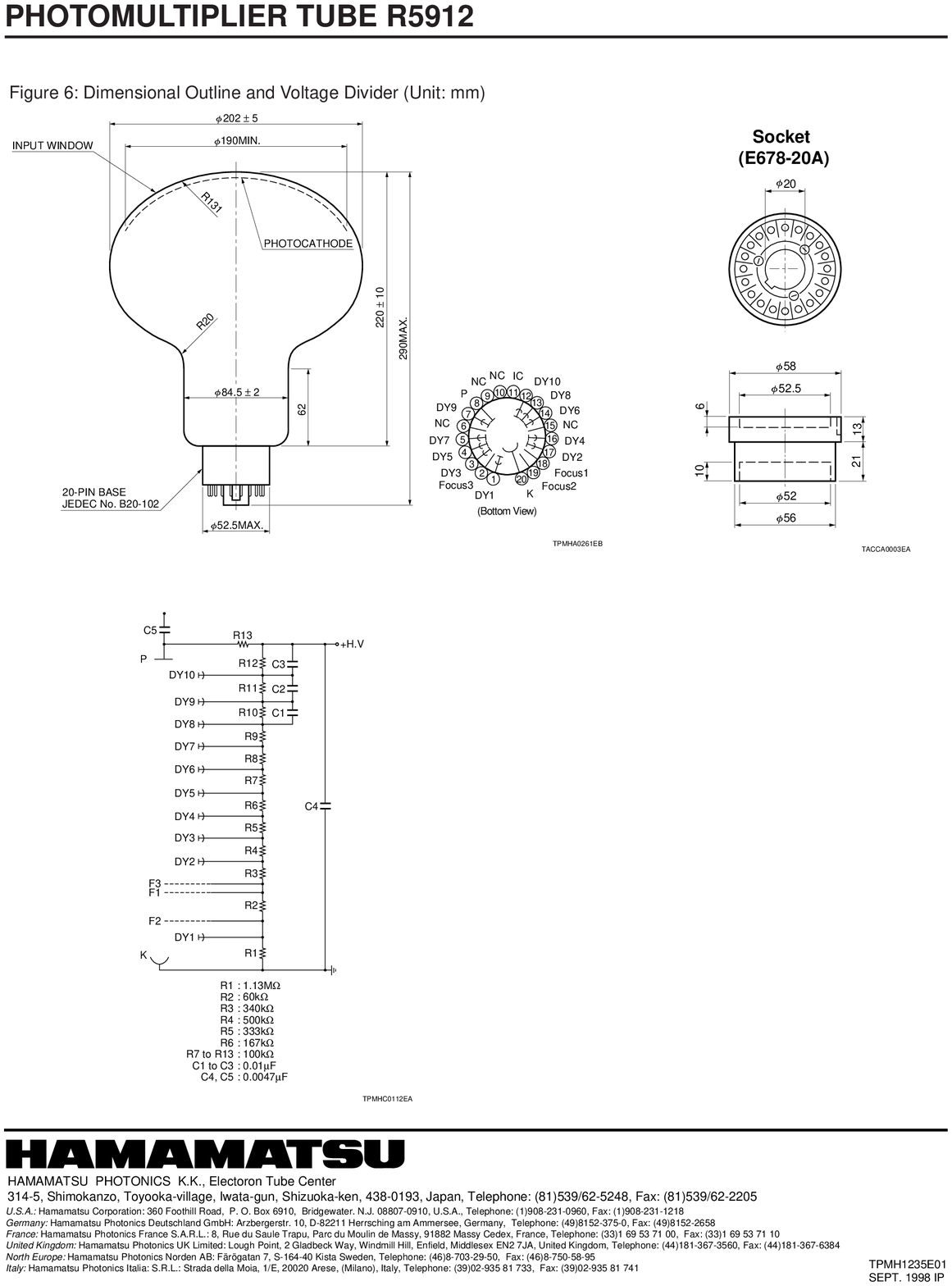}
\caption{Technical design specifications for the R5912 phototubes from Hamamatsu\cite{hamamatsu website}. Published with permission from Hamamatsu Photonics.}
\label{hamamatsu_drawing_dynode}
\end{figure}


\begin{thebibliography}{00}


% notes:

% \bibitem{label2}
% If there is a note, it should come last:
% \bibitem{label3} \note
% \end{subbibitems}

%\bibitem{}

%1\cite{booneproposal}
\bibitem{booneproposal}
Proposal for the MiniBooNE experiment:  http://www-boone.fnal.gov/publicpages/proposal.ps.

%2\bibitem{LSND}
%\cite{Athanassopoulos:1996jb}
\bibitem{Athanassopoulos:1996jb}
  C.~Athanassopoulos {\it et al.}  [LSND Collaboration],
  %``Evidence for anti-nu/mu $\to$ anti-nu/e oscillation from the LSND
  %experiment at the Los Alamos Meson Physics Facility,''
  Phys.\ Rev.\ Lett.\  {\bf 77}, 3082 (1996)
  [arXiv:nucl-ex/9605003].
  %%CITATION = NUCL-EX 9605003;%%

%3
\bibitem{hamamatsu website}
Hamamatsu Catalog, ``Large Photocathode Area PMTs", Catalog No.TPMH1286E02.\\
http://usa.hamamatsu.com/\\
http://www.hpk.co.jp/Eng/main.htm .

\bibitem{masterbond}
MASTER BOND INC. 154 Hobart Street, Hackensack, New Jersey 07601. \\
Phone: (201) 343-8983\\
Fax: (201) 343-2132\\
Web Site: http://www.masterbond.com

\bibitem{hysol}
Loctite brand product, 
http://www.loctite.com.\\
In US Toll Free: 800-LOCTITE (800-562-8483)\\
In Canada, call: 800-263-5043\\
Henkel Corp - Industrial, 
1001 Trout Brook Crossing,
Rocky Hill, CT 06067

%4\cite{Fleming:2002mv}
\bibitem{Fleming:2002mv}
  B.~T.~Fleming, L.~Bugel, E.~Hawker, V.~Sandberg, S.~Koutsoliotas, S.~McKenney and D.~Smith,
  %``Photomultiplier tube testing for the MiniBooNE experiment,''
  IEEE Trans.\ Nucl.\ Sci.\  {\bf 49}, 984 (2002).
  %%CITATION = IETNA,49,984;%%

%5
%\bibitem{dataq6} E.~Hawker, S.~McKenney, D.~Smith, and V.~Sandberg.
%\textsl{A data acquisition program written in C for the
%VXI crate.}  Aug-Sep.~ 2000.  %There is no reference for this program.  

%6
%\bibitem{filters} 

%7
\bibitem{Justin IEEE}
J.~May, M.~Wysocki, L.~Bugel, B.T.~Fleming, P.~Nienaber, and D.~Smith,
%Operational properties of photomultiplier tubes in the MiniBooNE experiment
Nuclear Science Symposium Conference Record, IEEE, {\bf 1}, 10-16 Nov.~2002, pp.~446-449, vol.~1.

%8
\bibitem{SNO}
M.T. Lyon,
%title = "Neutron Transport in the Sudbury Neutrino Detector",
 PhD thesis, University of Oxford 59 (2002).

%9
\bibitem{super-k} 
http://neutrino.phys.washington.edu/$\sim$superk.

%10
\bibitem{super-k news}
Fermi News, Vol. 24 no.19, Fri Nov. 23, 2001



\end{thebibliography}
\end{document}